\documentclass[accept, 12pt]{elsarticle}

\graphicspath{ {./figures/} }
\usepackage{hyperref}
\usepackage{float}
\usepackage{verbatim} 
\usepackage{apalike}
\usepackage{amsmath}
\usepackage{amssymb}
\usepackage{multirow}
\usepackage{graphicx}
\usepackage{tabularx}
\usepackage{color}
\usepackage{booktabs}
\usepackage{lscape}
\usepackage{placeins}
\usepackage{caption}
\usepackage{subcaption}
\restylefloat{figure}
\restylefloat{table}

\journal{Expert Systems with Applications}

\biboptions{authoryear}

\begin{document}

\begin{frontmatter}

\title{Constructing Time-Series Momentum Portfolios with Deep Multi-Task Learning}

\author[label1]{Joel Ong}
\ead{joel$\_$ong@mymail.sutd.edu.sg}

\author[label1]{Dorien Herremans}
\ead{dorien$\_$herremans@sutd.edu.sg}

\cortext[cor1]{Joel Ong}
\address[label1]{Singapore University of Technology and Design, 8 Somapah Rd, Singapore 487372}

\begin{abstract}
A diversified risk-adjusted time-series momentum (TSMOM) portfolio can deliver substantial abnormal returns and offer some degree of tail risk protection during extreme market events. The performance of existing TSMOM strategies, however, relies not only on the quality of the momentum signal but also on the efficacy of the volatility estimator. Yet many of the existing studies have always considered these two factors to be independent. Inspired by recent progress in Multi-Task Learning (MTL), we present a new approach using MTL in a deep neural network architecture that jointly learns portfolio construction and various auxiliary tasks related to volatility, such as forecasting realized volatility as measured by different volatility estimators. Through backtesting from January 2000 to December 2020 on a diversified portfolio of continuous futures contracts, we demonstrate that even after accounting for transaction costs of up to 3 basis points, our approach outperforms existing TSMOM strategies. Moreover, experiments confirm that adding auxiliary tasks indeed boosts the portfolio's performance. These findings demonstrate that MTL can be a powerful tool in finance.
\end{abstract}


\end{frontmatter}


\section{Introduction}
\label{introduction}

This section introduces the research background, literature review and research gaps, and our contribution, respectively. In Section \ref{sec:background}, we discussed the phenomenon of Momentum in Finance. Section \ref{sec:literature_review} reviews the literature associated with our study and identifies the current research gaps. Finally, section \ref{sec:contributions} summarizes our contributions.

\subsection{Background}
\label{sec:background}

Momentum is a phenomenon that has been extensively studied in the finance literature. It is based on the idea that assets that have recently outperformed their peers will continue to outperform their peers~\citep{jegadeesh_titman_1993,jegadeesh_titman_2001}. \citet{moskowitz_2012} proposed the concept of time-series momentum (TSMOM), whereby asset selection focuses purely on the assets’ past return rather than their relative returns. They show that a strategy where one simply buys assets if their past 12-month returns are positive and sells them otherwise can achieve an impressive risk-adjusted return. Other researchers have also investigated this phenomenon on a broader range of asset classes with extended sample periods and observed similar results~\citep{georgopoulou_wang_2016, levine_2016, hurst_2017}. Another critical aspect of the time-series momentum strategy is volatility scaling to target a constant volatility exposure. This involves leveraging exposure during low volatility periods and scaling back exposure during high volatility periods. This mechanism may reduce the probability of extreme losses by limiting the tail risk of extreme returns. Without such risk adjustments, momentum strategies are susceptible to large crashes during periods of market stress~\citep{barroso_2015, daniel_2016}. Furthermore,~\cite{baltas_2012} show that the efficacy of the volatility estimator can improve the performance of time-series momentum strategies. 

\subsection{Literature Review and Research gap}
\label{sec:literature_review}

The application of deep learning in finance is not new~\citep{francis_2001, cao_2003, kim_2003, sapankevych_2009, manuel_nunes_2019, law_2017, amine_2020, yu_2022, zou2022multimodal}. \cite{zhang_zohren_roberts_2020} proposes a deep learning approach to directly optimize the portfolio Sharpe ratio without requiring the expected return forecast. In addition, the authors utilize exchange-traded funds (ETFs) of market indices instead of individual assets to form a portfolio. Within the realm of momentum portfolio, \cite{bryan_2019} introduce a new trading strategy for time series momentum using deep learning techniques called Deep Momentum Networks (DMN). This approach combines deep learning-based trading rules with the volatility scaling framework of time series momentum to learn trend estimation and position sizing in a data-driven manner. The authors optimize the Sharpe ratio of the signal and backtest it on a portfolio of 88 continuous futures contracts. More recently, \cite{kieran_2021} introduce the Momentum Transformer, an attention-based deep-learning architecture~\citep{vaswani_2017} that outperforms benchmark time-series momentum strategies. Unlike LSTM architectures, the Momentum Transformer uses an attention mechanism to learn longer-term dependencies, adapt to new market regimes, and capture concurrent regimes at different timescales. 

In today's context, the return-generating and risk-management functions are typically separate entities. As a result, the portfolio construction process may need more oversight with adequate risk management before live implementation~\citep{harvey_rattray_hemert_2021}. In addition, much of the existing research on deep learning for portfolio construction focuses on maximizing the Sharpe ratio as an objective function. A yet unexplored type of type of deep learning in this context is Multi-task learning (MTL). Multi-task learning (MTL) involves training a neural network to learn multiple tasks (related to one another) jointly rather than independently to improve the model's generalizability~\citep{caruana_1997, thrun_1998, sebastian_2017}. This technique has become increasingly important due to its relevance in many applications, ranging from Computer vision, Natural Language Processing, to Speech Recognition~\citep{collobert_2008, bingel_2017, anastasopoulos_2018}. For example, when performing visual scene understanding in the field of Computer Vision, a model needs to understand both the scene's geometry and its semantics simultaneously. Not surprisingly, researchers showed that the MTL approach outperforms models trained individually on each task~\citep{cipolla_2018}. \citet{ghosn_1996} first used MTL to train a model to predict the future returns of multiple stocks simultaneously. However, the authors found significantly better results when some or all of the parameters of the stock models are not shared. Since then, much more research has been done on multi-task learning in finance, but much of this is focused on forecasting forward returns, ranking assets and not portfolio construction ~\citep{tao_ma_2022, yanzhe_2022, chenxun_2023}.

To address the research gaps mentioned above and inspired by recent progress in MTL, we propose constructing a time-series momentum portfolio through multi-task learning where tasks related to returns and risk are jointly learned. Our suggested approach envisions a union between the return-generating and risk-management processes. In this case, our auxiliary tasks, which focus on predicting multiple types of forward-looking volatility, allow the model to learn the shared representation of the assets' forward-looking risk relevant to the main tasks of portfolio construction. For instance, with a forward-looking view of the risk of an asset, we can size the asset's position more adequately. This approach can help capture complex interactions between investment objectives, such as risk and return trade-offs.

\subsection{Our Contributions}
\label{sec:contributions}

We demonstrate that using a multi-task deep neural architecture to learn portfolio construction and auxiliary tasks simultaneously improves the performance of time-series momentum strategies, thus making our approach more effective than existing TSMOM strategies. In addition, compared to the existing time-series momentum approach, our proposed approach circumvents the need to specify the momentum and position sizing rule explicitly. Our main contributions are: i.) A novel multi-task deep learning-based framework that results in improved time-series momentum portfolios, ii.) the first implementation and study of multi-task learning in the context of portfolio construction, and iii.) extensive experimental analysis to understand the performance trade-offs resulting from using different volatility-related auxiliary tasks.

The paper is organized as follows. In Section \ref{sec:classical}, we discuss existing work on time-series momentum strategies. Section \ref{sec:methodology} presents the proposed multi-task time-series momentum model. Section \ref{sec:setup} presents the setup of the different experiments, such as which dataset was used, benchmark models, and proposed backtesting strategy. In Section \ref{sec:results}, the results of our experiments are presented. Finally, in Section \ref{sec:conclusion} we summarize our findings and suggest directions for future research.

\section{Classical Time-Series Momentum Strategy}
\label{sec:classical}

One of the most influential ideas in the finance literature is the Efficient Market Hypothesis (EMH)~\citep{fama_1970}. In the most general sense, the EMH implies that any predictable pattern will eventually cease to exists as market participants increasingly act upon it. Unfortunately, the EHM fails to explain many of the persistent market anomalies documented by researchers~\citep{jegadeesh_titman_1993, kishore_2008, moskowitz_2012, rossi_2015}. One such market anomaly is cross-sectional momentum, whereby assets that recently outperformed their peers (over the past 3 to 12 months) will continue to outperform them on average over the next month~\citep{jegadeesh_titman_1993}. Since the study by~\citet{jegadeesh_titman_1993} was published, a growing body of research suggests that the performance of portfolios constructed using cross-sectional momentum is robust across asset classes, markets, and periods~\citep{jegadeesh_titman_2001, asness_2014, geczy_2016}.

In contrast to traditional work on momentum, \citet{moskowitz_2012} discovered another form of momentum: time-series momentum (TSMOM), whereby there is strong predictability from an asset's past performance. Specifically, they found that the past 12-months' excess return of an asset is a positive predictor of its future returns. Following \citet{moskowitz_2012}, we can express the realized returns $r_{t,t+1}^{TSMOM, i}$ for period $t$ to $t+1$ of a single asset $i$, for the TSMOM strategy as follows: 

\begin{equation} \label{eq:1}
r_{t,t+1}^{\text{TSMOM}, i} = sgn(r_{t - 252, t}^{i}) \frac{\sigma_{tgt}}{\sigma_{t}^{i}} r_{t,t+1}^{i}
\end{equation}

\begin{equation} \label{eq:2}
sgn(x) := \begin{cases} \mbox{-1} & \mbox{if } x < 0, \\ \mbox{0} & \mbox{if } x = 0, \\ \mbox{1} & \mbox{if } x > 0. \end{cases}
\end{equation}

\noindent where $r_{t - 252, t}^{i}$ is the previous one year return of asset $i$, $r_{t, t+ 1}^{i}$ is the one day return of the asset $i$, $\sigma_\text{tgt}$ is the  annualized target volatility, and $\sigma_{t}^{i}$ is the ex-ante volatility of asset $i$ estimated using the standard deviation of returns with exponentially decaying weights and a span of 60 days. In Equation~(\ref{eq:1}), the size of the trade is determined by $\sigma_{tgt} / \sigma_{t}^{i}$. Sizing the position in such a manner results in more volatile assets having a lower weight than their less volatile counterparts. A final (full) portfolio can be constructed using a TSMOM strategy by taking the average return series of the individual volatility scaled strategies where $r_{t, t+1}^{\text{TSMOM}}$ is the realised return of the portfolio strategy from day $t$ to $t + 1$ and $S_{t}$ is the total number of assets in the portfolio in period $t$, as per the equation below: 

\begin{equation}  \label{eq:3}
r_{t,t+1}^{\text{TSMOM}} = \frac{1}{S_{t}} \sum_{i=1}^{S_t} r_{t,t+1}^{\text{TSMOM}, i}
\end{equation}

\section{Methodology}
\label{sec:methodology}

\subsection{Multi-Task Time-Series Momentum Strategy}

\begin{figure}[ht!]
  \centering
  \includegraphics[width=0.9\linewidth]{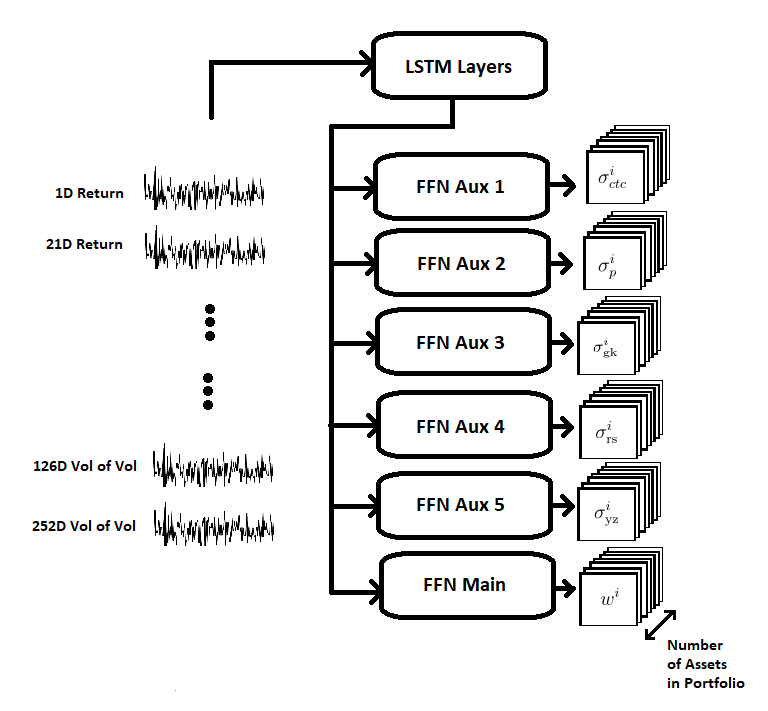}
  \caption{Proposed Multi-task learning architecture. The architecture consists of Long-Short Term Memory (LSTM) modules which serve as the shared layers. In these layers, the network parameters are shared across the various task-specific layers. Each task-specific layer is a feedforward network (FNN) layer, whereby each FNN learns a set of parameters optimized to their respective task. In Section \ref{sec:feature_set}, we discussed the features used as input to our model in more detail. In this section we will describe the output of each FNN.}
  \label{fig:mtl_lstm}
\end{figure}

In this work, we propose a deep-MTL architecture with hard parameter sharing, which includes two types of tasks: the main task and the auxiliary tasks. The main task of the proposed model is constructing a time-series momentum portfolio. The auxiliary tasks, on the other hand, are included to enhance the model's shared representation and allow the model to better generalize, which in turn will aid the main task. In our proposed architecture (Figure~\ref{fig:mtl_lstm}), the inputs are encoded using Long Short-Term Memory (LSTM)~\citep{schmidhuber_1997} layers which serve as the shared layers, after which the network branches out into task-specific feedforward networks. In Section \ref{sec:setup}, we describe the details of the architecture's implementation as well as the hyperparameters tuning process.

\subsection{Main Task: portfolio construction}

As per Equation~(\ref{eq:1}), the two key variables for building a traditional time-series momentum portfolio are the momentum signal used to identify trends and the volatility estimator used in position sizing. These are represented by the terms $sgn(r_{t-252, t}^{i})$ and $\sigma_{t}^{i}$ respectively. The product of the time-series momentum signal and the inverse ex-ante volatility for a particular asset will yield its resulting weight in the portfolio. Instead of explicitly defining time-series momentum signals and volatility estimators independently as proposed by \citet{moskowitz_2012}, we parameterize both of these terms and predict them directly using our proposed multi-task architecture, so as to learn both the trend and position sizing simultaneously. We can express the realized returns $r_{t,t+1}^{\rho}$ of our proposed MTL method as follows: 

\begin{equation} \label{eq:4}
r_{t,t+1}^{\rho} = \frac{\sigma_{tgt}}{S_{t}} \sum_{i=1}^{S_t} w_{t-1, t}^{i} \; r_{t,t+1}^{i}
\end{equation}

\noindent where $w_{t-1, t}^{i}$ is the output of the task-specific layer (for the main task) at time~$t$ for asset $i$ in our proposed model and represents the weight allocated to that asset $i$ in our portfolio. Regular re-balancing is typically required to manage the changing risk profile of each asset $i$ n any portfolio. Too many trades, however, could also hurt performance. To ensure that our model considers this and is thus suitable for live implementation, we embedded transaction costs into Equation (\ref{eq:4}) to obtain Equation (\ref{eq:5}) to calculate the realized returns more accurately. Here, $\tau$ is the transaction cost (set as 3 basis points (bps)), and $|w_{t-1, t}^{i} - w_{t-2, t-1}^{i}|$ is the change in weight of asset $s$ in the portfolio from time~$t$ to time $t+1$.

\begin{equation} \label{eq:5}
r_{t,t+1}^{\rho} = \frac{\sigma_{tgt}}{S_{t}} \sum_{i=1}^{S_t} w_{t-1, t}^{i} \; r_{t,t+1}^{i} - \tau |w_{t-1, t}^{i} - w_{t-2, t-1}^{i}|
\end{equation}

Since our primary objective when developing the multi-task model is to construct a portfolio with better risk-adjusted performance, we employ the Sharpe ratio~\citep{sharpe_1994}, as per Equation (\ref{eq:6}), as the loss function that is minimized during the training of our neural network model. 
\begin{equation} \label{eq:6}
L_{\text{Sharpe Ratio}} = -\frac{\mathbb{E}[r^{\rho}]}{\sigma_{r^{\rho}}}
\end{equation}

\noindent where $\mathbb{E}[r^{\rho}]$ and $\sigma_{r^{\rho}}$  are the mean and standard deviation of the portfolio's realised returns, respectively.

Unlike existing mean-variance approaches to portfolio construction~\citep{harry_1952, black_1992, chekhlov_2005, estrada_2007}, where forward-looking excess returns are required, our approach circumvents the need for this forecasting and obtains the asset allocations directly by optimizing the Sharpe ratio. As a result, we do not have to deal with instability caused by estimation errors when predicting excess returns~\citep{black_1992, clarke_2002, demiguel_2007} in our portfolio construction process.

In addition, having an optimized end-to-end model is extremely attractive as it circumvents the need to specify the time-series momentum signals and volatility estimator independently. Furthermore, using ex-ante volatility to scale the positions in the portfolio, as is the case for TSMOM, has its shortcomings. For example, the ex-post volatility of a time-series momentum portfolio could differ from the target volatility because the ex-ante volatility estimates are prone to estimation errors. Moreover, the volatility of the assets can change drastically during the holding period. These limitations may give rise to unnecessary turnover, reducing the performance after accounting for transaction costs~\citep{baz_2015b}.

\subsection{Auxiliary Tasks}

In the context of multi-task learning, the purpose of these auxiliary tasks is to enable the model to learn a shared representation that contains relevant information which may improve the performance of the main task~\citep{baxter_2000, sebastian_2017, lukas_2018}. As found by~\cite{baltas_2012}, the efficacy of the volatility estimator used to perform volatility scaling can significantly impact the performance of time-series momentum strategies. Hence, we study five auxiliary tasks for forecasting each asset's 21-days forward-volatility as defined by the:  

\begin{enumerate}
    \item Close-to-close volatility estimator, $\sigma_{ctc}$
    \item Parkinson volatility estimator, $\sigma_{p}$ \citep{parkinson_1980}
    \item Garman-Klass volatility estimator, $\sigma_{gk}$ \citep{garman_1980}
    \item Rogers-Satchell volatility estimator, $\sigma_{rs}$ \citep{rogers_1991}
    \item Yang-Zhang volatility estimator, $\sigma_{yz}$ \citep{yang_2000}
\end{enumerate}

By adding these auxiliary tasks, we aim to induce forward-looking volatility information into the shared representation, thus allowing the main task-specific layers to construct a better risk-adjusted time-series momentum portfolio. This is confirmed in the experiment in Section~\ref{experiment results and analysis}. As our training loss function for the auxiliary task-specific layers, we minimize the negative correlation between the predicted and the realized volatility:  

\begin{equation} \label{eq:7}
L_{\text{corr}(y, \hat{y})} = -\frac{S_{y, \hat{y}}}{S_{y} \cdot S_{\hat{y}}}
\end{equation}

\noindent where $y$ is the predicted volatility, $\hat{y}$ is the realized volatility, $S_{y, \hat{y}}$ is the covariance of $y$ and $\bar{y}$, which is calculated as $S_{y, \hat{y}} = \frac{1}{n-1}\sum_{i=1}^{n}(y_i - \bar{y})(\hat{y}_i - \bar{\hat{y}})$ where $n$ is the number of samples and $S_{y}$ and $S_{\hat{y}}$ is the standard deviation of $y$ and $\hat{y}$, respectively.

\subsection{Loss function}

Putting it all together, the final loss function of our model, which we minimize during training, can be written as Equation~(\ref{eq:8}). Here, the loss for the main task $L_{main}$ is the negative Sharpe ratio, and the loss for the $H$ auxiliary tasks $L_{aux, h}$ is the negative correlation between the predicted volatility and realized volatility. Finally, $\mu$ and $\lambda$ are the weights assigned to each of the tasks' loss, which are in the range of $[0,1]$ with $\mu + \lambda = 1$. Simply put, we perform a weighted linear sum of the losses for each task, where $\mu = 0.5$ and $\lambda = 0.5$. Here, $H$ is the set of five auxiliary tasks that include forecasting volatility defined by the different volatility estimators discussed above.

\begin{equation}  \label{eq:8}
\begin{split}
L_{total} &= \mu \; L_{\text{main}} + \lambda \sum_{h \in H} \; L_{\text{aux}, h}\\
          &= \mu \; L_{\text{Sharpe Ratio}} + \lambda \sum_{h \in H} \; L_{\rho_{y, \hat{y}}, h} \\
          &= -\mu \; \frac{\mathbb{E}[r^{\rho}]}{\sigma_{r^{\rho}}} + \lambda \sum_{h \in H} \;-\frac{S_{y_{h}, \hat{y}_{h}}}{S_{y_{h}} \cdot S_{\hat{y}_{h}}}
\end{split}
\end{equation}

\section{Experimental Setup}
\label{sec:setup}

\subsection{Dataset}

Individual futures contracts have expiry dates and variable liquidity. As a result, they are unsuitable for analyzing long-term trends. Instead, we evaluate our proposed model on the Stevens Continuous Futures data feed~\citep{steven_analytics} obtained through Nasdaq Data Link from January 1990 to December 2020. This data feed has a daily frequency and more than 20 years of historical data. The Steven Continuous Futures data feed provides a long-term continuous price history for 78 of the most popular US and international futures contracts from various asset classes, including commodities, currencies, fixed income, and equity indexes futures. In particular, we use the continuous contract history of each asset spliced together using end-to-end concatenation with the prices adjusted using the backward-ratio method.

\subsection{Feature Set}
\label{sec:feature_set}

We derive a set of time-series momentum features from the daily settled price of the continuous futures by taking the log returns ($r_{t - d, t}^{i}$) over the past~1 trading day, 21 trading days, 63 trading days, 126 trading days, and finally 252 trading days: 

\begin{equation} \label{eq:9}
r_{t - d, t}^{i} = \ln{\frac{P_t^i}{P_{t - d}^i}}
\end{equation}

\noindent where $r_{t - d, t}^{i}$ is the natural logarithm of the $d$-day return of asset $i$ at day $t$, $P_{t}^{i}$ is the settled price of asset $i$ at time $t$ and $P_{t-d}^{i}$ is the settled price of asset $i$, $d$ trading days ago at time $t$. We also constructed features that would allow the proposed model to capture various asset risk dimensions:

\begin{enumerate}

  \item Realized volatility $RV_{t-n}^{i}$ over the past 5 trading days, 21 trading days, 63 trading days, 126 trading days, and finally 252 trading days. The realized volatility of an asset $i$ is given by Equation \ref{eq:10}.

\begin{equation} \label{eq:10}
RV_{t-n, t}^{i} = \sqrt{\frac{252}{N}\sum_{n=1}^{N} (r_{t-n, t-n+1}^{i})^2}
\end{equation}

\noindent where $RV_{t-n}^{i}$ is the realized volatility of asset $i$ at time $t$. The constant 252 represents the approximate number of trading days in a year, $N$ is the number of trading days in the measurement time frame, $r_{t-n, t-n+1}^{i}$ is the log daily return of asset $i$ at time $t - n$.

  \item Realized volatility of realized volatility (vol of vol) is obtained by applying the same formula used for calculating the realized volatility on the realized volatility itself. When applying the formula the second time, however, we always calculate it on a 21 trading day basis, regardless of the time frame of the realized volatility.

\end{enumerate}

There are two reasons why we created the features as such. First, to retain the spirit of time-series momentum, we only use features similar to those used in constructing a time-series momentum portfolio~\citep{moskowitz_2012}. Second, and most importantly, we do not want the feature engineering to drive the performance of the portfolios, but rather the model's multi-task architecture should be delivering the performance.

Given that these features have varying scales, before using them as inputs to our model, we apply z-score standardization with a sliding window of length 21 to our features. In this case, a sliding window z-score is preferred over a static z-score because financial time-series data are non-stationary.

\subsection{Benchmark Models}

We have constructed a number of portfolios to serve as the reference benchmarks for our proposed model's performance.

\begin{enumerate}
  \item TSMOM is the time-series momentum portfolio by \citet{moskowitz_2012}, whereby assets are bought or sold based on the assets' previous 12 months return. This portfolio is re-balanced daily.
  
  \item CTA-MOM is the Commodity Trading Advisor (CTA)-momentum signal based on the cross-over of exponentially weighted moving averages as proposed by \citet{baz_2015b}. This portfolio is re-balanced daily.

\end{enumerate}

The volatility target, $\sigma_\text{tgt}$ for TSMOM and CTA-MOM was chosen such that the average return across all assets in the portfolios would have approximately 10\% annualized volatility from January 2000 to December 2020.

\subsection{Backtest Specifications}

\begin{figure}[ht!]
  \centering
  \includegraphics[width=\linewidth]{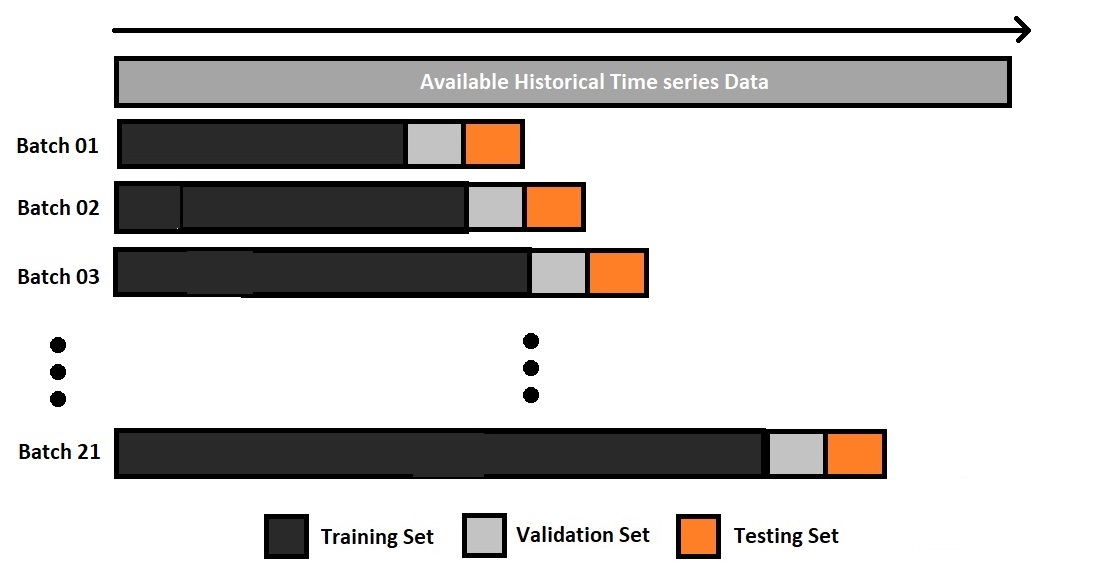}
  \caption{Expanding window cross-validation.}
  \label{fig:expanding_window}
\end{figure}

In the backtest results presented in the next section, our proposed model was trained from scratch every year using an expanding window cross-validation approach as shown in Figure~\ref{fig:expanding_window}, with 20\% of the training data kept as a validation set. After training, we used the models to generate the portfolio on the test set, with each test set containing a year's worth of out-of-sample results. Then, using data from January 1990 to December 2020, we repeated this expanding window out-of-sample generation 21 times, resulting in an out-of-sample backtest from January 2000 to December 2020.

\begin{table}[htb]
\centering
\caption{Hyperparameter Search Space.}
\begin{tabular}{lr}
\toprule

Parameters & Values\\
\midrule
Number of LSTM Layers & 1, 2, 3, 4 \\ 

LSTM Number of Hidden Units &  32, 64, 126, 252 \\ 

LSTM Dropout Rate & 0.05, 0.10, 0.15, 0.20 \\

Number of MLP Layers &  1, 2, 3, 4 \\ 

MLP Number of Hidden Units &  32, 64, 126, 252 \\ 

MLP Dropout Rate & 0.05, 0.10, 0.15, 0.20 \\

Learning Rate & 0.0001, 0.001, 0.01, 0.1 \\

Max Gradient Norm &  0.01, 0.1, 0. \\

\bottomrule
\end{tabular}
\label{table:1}
\end{table}

Throughout the backtest, we trained our model on the training data. During training, the loss functions (both the negative Sharpe loss function for the main-task and negative correlation for the auxiliary tasks) are minimized using Stochastic Gradient Descent (SGD) with the Adam optimizer and a learning rate set at $0.0001$. Finally, the validation set was used to perform a grid search in order to find the best hyperparameters within the parameter search space shown in Table \ref{table:1}.  The model training will terminate when it reaches 200 epochs. In addition, we implemented early stopping, which terminates training before the 200 epochs are reached when there is no longer an increase in the validation loss (on the validation set) for 25 epochs. Since we took an expanding window out-of-sample strategy to train and validate our model, the final training batch, which uses data from January 1990 to December 2019, took approximately 2 hours to train using a machine with NVIDIA GeForce RTX 2060 installed.

\section{Experiment results and analysis}
\label{sec:results}
\label{experiment results and analysis}

\subsection{Performance Evaluation}
\label{sec:performance_evaluation}

We compute the backtest metrics for all of the portfolios by aggregating the out-of-sample results from January 2000 to December 2020. We then derive the net backtesting metrics by assuming a transaction cost of 3 basis points (bps). During our backtest, we applied volatility scaling to all portfolios, bringing the volatility to a target of 10\%. This rescaling also helps to enable comparisons between the cumulative returns of different strategies. In Table \ref{table:2}, the returns of our proposed strategy, which we refer to as MTL-TSMOM, outperforms both TSMOM and CTA-MOM by 236 bps and 624 bps, respectively, over the 20 year test period.

\begin{figure}[htpb]
  \centering
  \includegraphics[width=\linewidth]{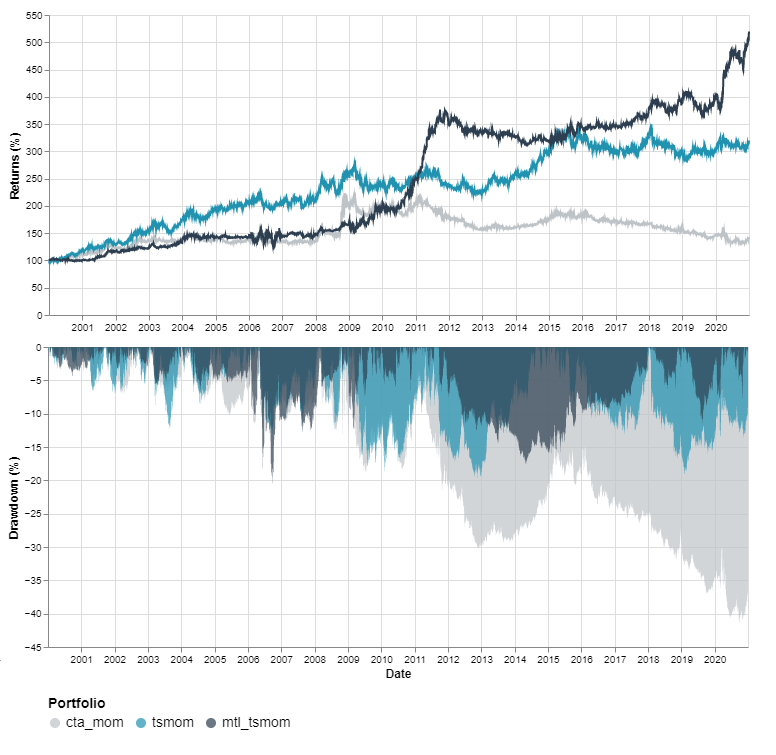}
  \caption{(Top) The plot shows our proposed model performance (dark blue) against the benchmark models over the test set period January 2000 to December 2020. At inception, we invested \$100 in each strategy, and the terminal value indicates how much our initial investment would be worth at the end of December 2020. All strategies were re-scaled to 10\% volatility. (Bottom) The plot shows the drawdown of our proposed model against the benchmark models during the same period. Drawdown refers to the decline in the value of a portfolio from a previous peak value to the largest subsequent trough.}
  \label{fig:strategies_cum_returns}
\end{figure}

\begin{table}[hbtp]
\caption{Backtest metrics (net) from January 2000 to December 2020. Transaction costs were set at 3 bps. All portfolios have been rescaled to target 10\% annualized volatility.}
\resizebox{\textwidth}{!}{%
\begin{tabular}{lrrr}
\toprule
 & \textbf{TSMOM} & \textbf{CTA-MOM} & \textbf{MTL-TSMOM} \\
 \midrule

Annualized Return (\%) & 5.54 &  1.66 & \textbf{7.90}\\ 

Annualized Sharpe Ratio & 0.59 &  0.21 & \textbf{0.81}\\ 

Annualized Sortino Ratio & 0.83 &  0.31 & \textbf{1.20}\\ 

Return Over Max Drawdown & 0.28 & 0.04 & \textbf{0.38} \\

Max Drawdown (\%) & \textbf{-19.88} & -41.63 & -21.00\\

Max Drawdown Period & 432 & 3066 & \textbf{103}\\

Max Drawdown Recovery Period & 385 & N.A. & \textbf{42}\\

Proportion of Positive Returns (\%) & \textbf{53.44} & 51.44 & 51.62\\
\bottomrule
\end{tabular}
\label{table:2}}
\end{table}

The maximum drawdown is a strategy's maximum loss from peak to trough over a given investment period. Drawdown presents a significant risk to investors when considering the uptick in returns needed to overcome a drawdown. For instance, if a strategy losses 50\%, it requires a 100\% increase to recover to the former peak. Throughout the backtest, our proposed model's maximum drawdown is -21\%, significantly better than the CTA-MOM strategy, which has a -41.63\% maximum drawdown but slightly worse than the TSMOM strategy, a maximum drawdown of -19.88\%. However, when considering the strategy's return divided by the maximum drawdown, our proposed model performs better with a score of 0.38. In contrast, the TSMOM and CTA-MOM benchmark models obtained a score of 0.28 and 0.04, respectively. In addition to a low drawdown magnitude, investors also typically prefer a portfolio with a lower recovery time. Hence, the drawdown duration is another critical dimension of analyzing a trading strategy's risk. This work presents two metrics to assess our proposed and benchmark models. First, the Max Drawdown Period is the total time (in trading days) that has elapsed from the start to the end of the most significant drawdown. Secondly, the Max Drawdown Recovery Period represents the time taken in trading days by the strategy (if recovered), from the largest drawdown to recovery. In Table~\ref{table:1}, our proposed model performs better than the benchmark models in terms of the Max Drawdown Period and Max Drawdown Recovery Period. Our proposed model began its largest drawdown period on 5 May 2006 and hit its max drawdown of -21\% on 2 October 2006. Subsequently, it took only 42 trading days to recover from its biggest drawdown. In comparison, TSMOM hit its maximum drawdown of -19.9\% on 28 December 2012 and took 385 trading days before making new highs, while CTA-TSMOM did not recover from its maximum drawdown in the backtest period. Although our proposed model did not fare better than TSMOM in terms of the magnitude of the drawdown, the lower drawdown recovery periods and higher annualized return present itself as a more attractive portfolio.

\subsection{Ablation Study}

Our proposed Multi-Task TSMOM model is trained with several auxiliary tasks. In this section, we perform an ablation study to investigate the effects of training the portfolio construction tasks with different auxiliary tasks.

\begin{table}[htb]
\caption{Backtest metrics (net) from January 2000 to December 2020 when training our model with the main task (portfolio construction) with various single auxiliary tasks (all different types of volatility forecasts), as well as a full model with all auxiliary tasks and one without any auxiliary tasks. The transaction costs were set at 3 bps. All portfolios have been rescaled to target 10\% annualized volatility.}
\label{table:3}
\resizebox{\textwidth}{!}{%
\begin{tabular}{lcccc}
\toprule
\textbf{Aux. task} & \textbf{Ann. Return (\%)} & \textbf{Sharpe Ratio} & \textbf{Sortino Ratio} & \textbf{Max Drawdown (\%)} \\
 \midrule
None & 6.63 & 0.69 & 1.02 & -36.54 \\
Close-to-close & 6.85 & 0.71 & 1.05 & -21.30 \\
Parkinson  & 4.61 & 0.50 & 0.74 & -38.20 \\
Garman Klas  & 6.67 & 0.70 & 1.03 & -34.82 \\
Rogers-Satchell  & 5.32 & 0.57 & 0.85 & \textbf{-18.69} \\
Yang-Zhang  & 7.03 & 0.73 & 1.08 & -25.66 \\
MTL-TSMOM (All) & \textbf{7.90} & \textbf{0.81} & \textbf{1.20} & -21.00\\
\bottomrule
\end{tabular}%
}
\end{table}

Table \ref{table:3} shows the risk-adjusted performance of the various portfolios constructed by models without any auxiliary task, with a single auxiliary task, and with all auxiliary tasks (MTL-TSMOM). The table allows us to compare the full proposed MTL-TSMOM to a model trained without any auxiliary tasks. The former MTL-TSMOM outperforms the latter by 127 bps and incurred a much lower max drawdown (-21\%). This better performance when using all auxiliary tasks is also apparent when comparing the full MTL-TSMOM to models where the portfolio construction task was co-trained with a single auxiliary task. This confirms the advantage of co-training the portfolio construction task with many auxiliary tasks in an MTL setting. We also note, however, that not all auxiliary tasks, when trained jointly with the portfolio construction task, results in better performance. These observations highlight the intricacies of task selection for portfolio construction in an MTL setting, and confirm that our proposed model with all five auxiliary tasks largely achieves the best performance. 

\subsection{Long-term Correlation with Equities Index}

It is widely known that the Time-Series Momentum Strategy (TSMOM) and Trend following strategies, such as CTA-MOM, exhibit negative downside correlation to equity markets \citep{hurst_2017}, providing the potential to perform well during periods of sustained stress in global equity markets. In this section, we show that our proposed model exhibits similar characteristics while attaining a lower long-term correlation to equity markets. We have selected the US MSCI Index as the equity index benchmark for our analysis. The US MSCI Index measures the performance of the large and mid-cap segments of the US market. It contains 626 constituents and covers approximately 85\% of the free float-adjusted market capitalization in the US. Instead of using the US MSCI Price Return Index for our analysis, we used the Total Return Index. The total return index accounts for activity associated with dividends, interest, rights offerings, and other distributions realized over a given period. As a result, the US MSCI Total Return Index is a better representation of the index's actual growth over time. Using the Bloomberg Terminal, we obtained the US MSCI Total Return Index data from January 2000 to December 2020.

\begin{figure}[hbtp]
     \centering
     \caption{Rolling Correlation of Momentum Portfolios against US MSCI Index}
    \label{fig:rolling_correlation_mom}
     \begin{subfigure}[b]{\linewidth}
         \centering
         \includegraphics[width=\linewidth]{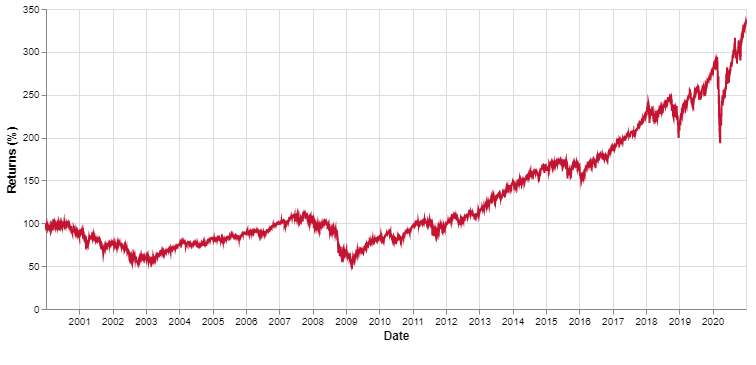}
         \caption{Cumulative Returns of US MSCI Total Return Index}
         \label{fig:total_return_index}
     \end{subfigure}
     \hfill
     \begin{subfigure}[b]{\linewidth}
         \centering
         \includegraphics[width=\linewidth]{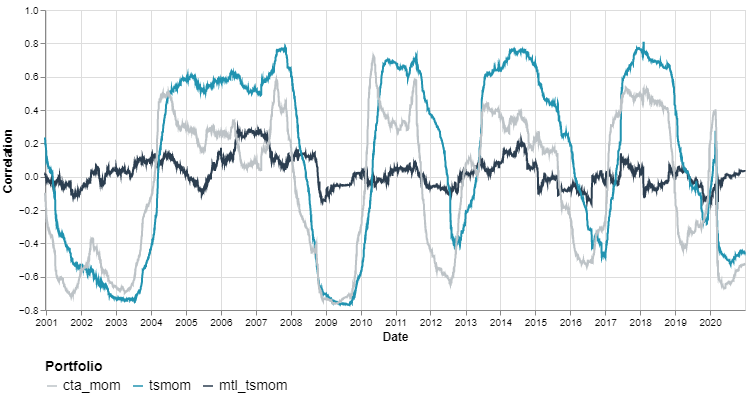}
         \caption{The plot shows the 252-trading day rolling correlation of our proposed model and benchmark momentum models' daily returns with the US MSCI Total Return Index returns.}
         \label{fig:rolling_correlation}
     \end{subfigure}
\end{figure}

\begin{figure}
     \centering
    \caption{Crisis Performance}
    \label{fig:crisis_performance}
     \begin{subfigure}[b]{\linewidth}
         \centering
         \includegraphics[width=\linewidth]{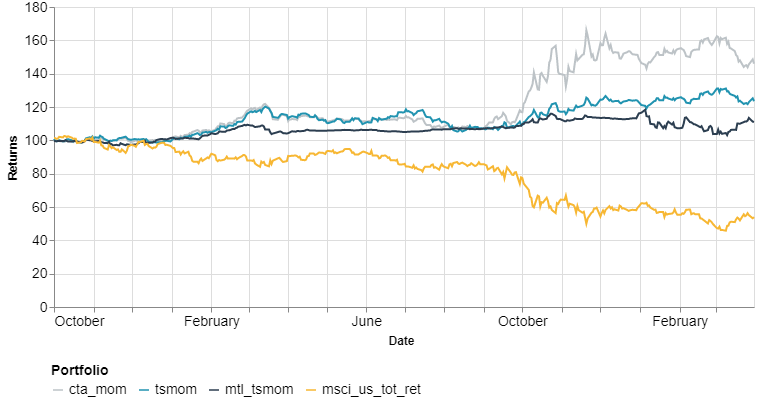}
         \caption{The plots show that our proposed strategy, two benchmark strategies, as well as the equity index performance during the Global Financial Crisis.}
         \label{fig:gfc_performance}
     \end{subfigure}
     \hfill
     \begin{subfigure}[b]{\linewidth}
         \centering
         \includegraphics[width=\linewidth]{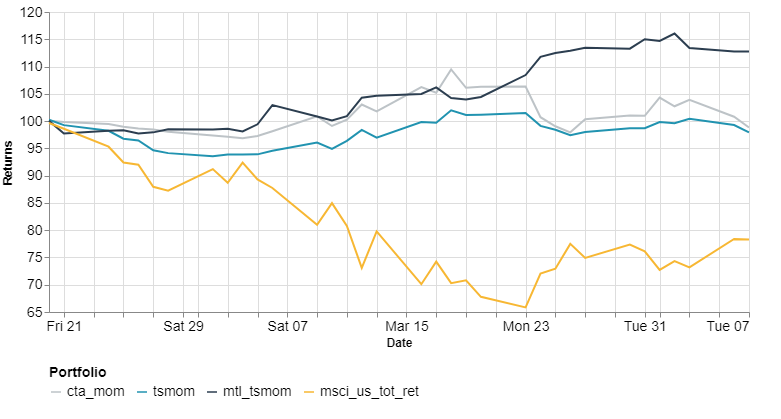}
         \caption{The plots show that our proposed strategy, two benchmark strategies, as well as the equity index performance during the Coronavirus Crash.}
         \label{fig:covid_performance}
     \end{subfigure}
\end{figure}

In Figure \ref{fig:rolling_correlation}, we plot the 252-day rolling correlation of the results generated by our proposed model in Section \ref{sec:performance_evaluation} as well as the two benchmark portfolios returns with the US MSCI Total Return Index, which measures the performance of the large and mid cap segments of the US market. From the figure, it can be observed that there are some crucial times when the negative correlation between all three models and the US MSCI Total Return Index coincides with periods of sustained negative equity performance. For example, we observe a dip in the 252-trading days rolling correlation, which becomes highly negative during periods of financial market stress, such as the Global Financial Crisis of 2008-2009~\citep{wood_giegerich_roberts_zohren_2022} and, more recently, the Coronavirus Crash of 2020~\citep{mazur_2021}. The longest bull market on record post-global financial crisis began in March 2009 and lasted nearly 11 years until the COVID-19 pandemic brought it to a close. We observed a relatively higher correlation between the benchmark models and US MSCI Total Return Index during this period. Interestingly, our proposed models maintain a relatively low correlation to the US MSCI Total Return Index throughout our backtesting period, ranging from -0.24 to 0.28 as compared to the benchmark models, where the 252-trading days rolling correlation is in the more extreme range of -0.77 to 0.81. The low long-term correlation with equities makes our proposed model extremely attractive for asset allocators looking for an alternative source of return that has a low correlation with the standard 60/40 portfolio~\citep{robinson_langley_2017, robertson_management_2020}.

To further substantiate the diversification benefits of our proposed model, we picked two well-known market stress events to observe how our proposed and benchmark models perform: the Global Financial Crisis (GFC) of 2008, which began in early October 2007 before bottoming in March 2009, and more recently, the Coronavirus Crash of 2020, which began in mid-February 2020 and ended in early April 2020. In Figure \ref{fig:crisis_performance}, we plot the cumulative return of our proposed strategy, two benchmark strategies, and the equity index during the GFC and Coronavirus Crash. During both of these periods of extreme market stress, our proposed and the benchmark models significantly outperformed the US MSCI Total Return Index, with our proposed model outperforming the US MSCI Total Return Index by 57\% and 34.4\% in the GFC and Coronavirus Crash of 2020, respectively.

\section{Conclusion}
\label{sec:conclusion}

We propose a multi-task learning time-series momentum model (MTL-TSM), which is a novel portfolio optimization method based on deep multi-task learning. In the MTL setting, our model jointly learns portfolio construction and various auxiliary tasks related to volatility, such as forecasting realized volatility. Our model was implemented in Python with PyTorch, the full source code of the model is available online\footnote{\url{https://github.com/joelowj/mtl-tsmom}}. In a thorough experiment, after accounting for transaction costs of up to 3 basis points, we demonstrate that our proposed approach consistently outperforms all benchmarks in risk-adjusted return, achieving an out-of-sample annualized return of 7.90\% for the period of 1st January 2000 to 31st December 2020. Furthermore, our extensive experiments also show that the proposed MTL-TSMOM outperforms existing multi-task benchmarks in constructing a better risk-adjusted portfolio. Lastly, our proposed model shows a low correlation to the US MSCI Total Return Index's returns, thus presenting itself as an extremely attractive diversifier to existing stocks and bonds portfolios. Future research directions include identifying additional auxiliary tasks to construct a time-series momentum portfolio with better overall performance. In addition, it would be of interest to explore dynamic task weighting instead of static task weighting. Finally, it would be interesting to quantify the relationship between portfolio construction and other auxiliary tasks. In particular, how valuable they may be for multi-task representation learning in portfolio construction.

\bibliography{sample}

\begin{thebibliography}{59}
\expandafter\ifx\csname natexlab\endcsname\relax\def\natexlab#1{#1}\fi
\providecommand{\url}[1]{\texttt{#1}}
\providecommand{\href}[2]{#2}
\providecommand{\path}[1]{#1}
\providecommand{\DOIprefix}{doi:}
\providecommand{\ArXivprefix}{arXiv:}
\providecommand{\URLprefix}{URL: }
\providecommand{\Pubmedprefix}{pmid:}
\providecommand{\doi}[1]{\href{http://dx.doi.org/#1}{\path{#1}}}
\providecommand{\Pubmed}[1]{\href{pmid:#1}{\path{#1}}}
\providecommand{\bibinfo}[2]{#2}
\ifx\xfnm\relax \def\xfnm[#1]{\unskip,\space#1}\fi
\bibitem[{Aboussalah \& Lee(2020)}]{amine_2020}
\bibinfo{author}{Aboussalah, A.~M.}, \& \bibinfo{author}{Lee, C.-G.}
  (\bibinfo{year}{2020}).
\newblock \bibinfo{title}{Continuous control with stacked deep dynamic
  recurrent reinforcement learning for portfolio optimization}.
\newblock {\it \bibinfo{journal}{Expert Systems with Applications}\/},  {\it
  \bibinfo{volume}{140}\/}, \bibinfo{pages}{112891}. \URLprefix
  \url{https://www.sciencedirect.com/science/article/pii/S0957417419306074}.
  \DOIprefix\doi{https://doi.org/10.1016/j.eswa.2019.112891}.
\bibitem[{Anastasopoulos \& Chiang(2018)}]{anastasopoulos_2018}
\bibinfo{author}{Anastasopoulos, A.}, \& \bibinfo{author}{Chiang, D.}
  (\bibinfo{year}{2018}).
\newblock \bibinfo{title}{Tied multitask learning for neural speech
  translation}.
\newblock {\it \bibinfo{journal}{Proceedings of the 2018 Conference of the
  North {A}merican Chapter of the Association for Computational Linguistics:
  Human Language Technologies, Volume 1 (Long Papers)}\/},  (pp.
  \bibinfo{pages}{82--91}).
\bibitem[{Asness et~al.(2014)Asness, Frazzini, Israel \&
  Moskowitz}]{asness_2014}
\bibinfo{author}{Asness, C.}, \bibinfo{author}{Frazzini, A.},
  \bibinfo{author}{Israel, R.}, \& \bibinfo{author}{Moskowitz, T.}
  (\bibinfo{year}{2014}).
\newblock \bibinfo{title}{Fact, fiction, and momentum investing}.
\newblock {\it \bibinfo{journal}{The Journal of Portfolio Management}\/},  {\it
  \bibinfo{volume}{40}\/}, \bibinfo{pages}{75--92}.
\bibitem[{Baltas \& Kosowski(2012)}]{baltas_2012}
\bibinfo{author}{Baltas, A.}, \& \bibinfo{author}{Kosowski, R.}
  (\bibinfo{year}{2012}).
\newblock \bibinfo{title}{Improving time-series momentum strategies: The role
  of trading signals and volatility estimators}.
\newblock {\it \bibinfo{journal}{SSRN Electronic Journal}\/}, .
\bibitem[{Barroso \& Santa-Clara(2015)}]{barroso_2015}
\bibinfo{author}{Barroso, P.}, \& \bibinfo{author}{Santa-Clara, P.}
  (\bibinfo{year}{2015}).
\newblock \bibinfo{title}{Momentum has its moments}.
\newblock {\it \bibinfo{journal}{Journal of Financial Economics}\/},  {\it
  \bibinfo{volume}{116}\/}, \bibinfo{pages}{111--120}.
\bibitem[{Baxter(2000)}]{baxter_2000}
\bibinfo{author}{Baxter, J.} (\bibinfo{year}{2000}).
\newblock \bibinfo{title}{A model of inductive bias learning}.
\newblock {\it \bibinfo{journal}{J. Artif. Int. Res.}\/},  {\it
  \bibinfo{volume}{12}\/}, \bibinfo{pages}{149–198}.
\bibitem[{Baz et~al.(2015)Baz, Granger, Harvey, Roux \& Rattray}]{baz_2015b}
\bibinfo{author}{Baz, J.}, \bibinfo{author}{Granger, N.},
  \bibinfo{author}{Harvey, C.}, \bibinfo{author}{Roux, N.}, \&
  \bibinfo{author}{Rattray, S.} (\bibinfo{year}{2015}).
\newblock \bibinfo{title}{Dissecting investment strategies in the cross section
  and time series}.
\newblock {\it \bibinfo{journal}{SSRN Electronic Journal}\/}, .
  \DOIprefix\doi{10.2139/ssrn.2695101}.
\bibitem[{Bingel \& S{\o}gaard(2017)}]{bingel_2017}
\bibinfo{author}{Bingel, J.}, \& \bibinfo{author}{S{\o}gaard, A.}
  (\bibinfo{year}{2017}).
\newblock \bibinfo{title}{Identifying beneficial task relations for multi-task
  learning in deep neural networks}.
\newblock {\it \bibinfo{journal}{Proceedings of the 15th Conference of the
  {E}uropean Chapter of the Association for Computational Linguistics: Volume
  2, Short Papers}\/},  {\it \bibinfo{volume}{2}\/}, \bibinfo{pages}{164--169}.
\bibitem[{Black \& Litterman(1992)}]{black_1992}
\bibinfo{author}{Black, F.}, \& \bibinfo{author}{Litterman, R.}
  (\bibinfo{year}{1992}).
\newblock \bibinfo{title}{Global portfolio optimization}.
\newblock {\it \bibinfo{journal}{Financial Analysts Journal}\/},  {\it
  \bibinfo{volume}{48}\/}, \bibinfo{pages}{28--43}.
  \DOIprefix\doi{10.2469/faj.v48.n5.28}.
\bibitem[{Cao \& Tay(2003)}]{cao_2003}
\bibinfo{author}{Cao, L.}, \& \bibinfo{author}{Tay, F.} (\bibinfo{year}{2003}).
\newblock \bibinfo{title}{Support vector machine with adaptive parameters in
  financial time series forecasting}.
\newblock {\it \bibinfo{journal}{IEEE Transactions on Neural Networks}\/},
  {\it \bibinfo{volume}{14}\/}, \bibinfo{pages}{1506--1518}.
  \DOIprefix\doi{10.1109/TNN.2003.820556}.
\bibitem[{Caruana(1997)}]{caruana_1997}
\bibinfo{author}{Caruana, R.} (\bibinfo{year}{1997}).
\newblock \bibinfo{title}{Multitask learning}.
\newblock {\it \bibinfo{journal}{Machine Learning}\/},  {\it
  \bibinfo{volume}{28}\/}, \bibinfo{pages}{41--75}.
\bibitem[{Chekhlov et~al.(2005)Chekhlov, Uryasev \& ZABARANKIN}]{chekhlov_2005}
\bibinfo{author}{Chekhlov, A.}, \bibinfo{author}{Uryasev, S.}, \&
  \bibinfo{author}{ZABARANKIN, M.} (\bibinfo{year}{2005}).
\newblock \bibinfo{title}{Drawdown measure in portfolio optimization}.
\newblock {\it \bibinfo{journal}{International Journal of Theoretical and
  Applied Finance (IJTAF)}\/},  {\it \bibinfo{volume}{08}\/},
  \bibinfo{pages}{13--58}. \DOIprefix\doi{10.2139/ssrn.544742}.
\bibitem[{Cipolla et~al.(2008)Cipolla, Gal \& Kendall}]{cipolla_2018}
\bibinfo{author}{Cipolla, R.}, \bibinfo{author}{Gal, Y.}, \&
  \bibinfo{author}{Kendall, A.} (\bibinfo{year}{2008}).
\newblock \bibinfo{title}{Multi-task learning using uncertainty to weigh losses
  for scene geometry and semantics}.
\newblock {\it \bibinfo{journal}{2018 IEEE/CVF Conference on Computer Vision
  and Pattern Recognition}\/}, .
\bibitem[{Clarke et~al.(2002)Clarke, de~Silva \& Thorley}]{clarke_2002}
\bibinfo{author}{Clarke, R.}, \bibinfo{author}{de~Silva, H.}, \&
  \bibinfo{author}{Thorley, S.} (\bibinfo{year}{2002}).
\newblock \bibinfo{title}{Portfolio constraints and the fundamental law of
  active management}.
\newblock {\it \bibinfo{journal}{Financial Analysts Journal}\/},  {\it
  \bibinfo{volume}{58}\/}, \bibinfo{pages}{48--66}.
  \DOIprefix\doi{10.2469/faj.v58.n5.2468}.
\bibitem[{Collobert \& Weston(2008)}]{collobert_2008}
\bibinfo{author}{Collobert, R.}, \& \bibinfo{author}{Weston, J.}
  (\bibinfo{year}{2008}).
\newblock \bibinfo{title}{A unified architecture for natural language
  processing: deep neural networks with multitask learning}.
\newblock {\it \bibinfo{journal}{Proceedings of the 25th international
  conference on Machine learning - ICML '08}\/},  {\it \bibinfo{volume}{28}\/},
  \bibinfo{pages}{41--75}.
\bibitem[{Daniel \& Moskowitz(2016)}]{daniel_2016}
\bibinfo{author}{Daniel, K.}, \& \bibinfo{author}{Moskowitz, T.~J.}
  (\bibinfo{year}{2016}).
\newblock \bibinfo{title}{Momentum crashes}.
\newblock {\it \bibinfo{journal}{Journal of Financial Economics}\/},  {\it
  \bibinfo{volume}{122}\/}, \bibinfo{pages}{221--247}.
\bibitem[{DeMiguel et~al.(2007)DeMiguel, Garlappi \& Uppal}]{demiguel_2007}
\bibinfo{author}{DeMiguel, V.}, \bibinfo{author}{Garlappi, L.}, \&
  \bibinfo{author}{Uppal, R.} (\bibinfo{year}{2007}).
\newblock \bibinfo{title}{{Optimal Versus Naive Diversification: How
  Inefficient is the 1/N Portfolio Strategy?}}
\newblock {\it \bibinfo{journal}{The Review of Financial Studies}\/},  {\it
  \bibinfo{volume}{22}\/}, \bibinfo{pages}{1915--1953}.
  \DOIprefix\doi{10.1093/rfs/hhm075}.
\bibitem[{Estrada(2007)}]{estrada_2007}
\bibinfo{author}{Estrada, J.} (\bibinfo{year}{2007}).
\newblock \bibinfo{title}{Mean-semivariance optimization: A heuristic
  approach}.
\newblock {\it \bibinfo{journal}{Journal of Applied Finance}\/},  {\it
  \bibinfo{volume}{18}\/}. \DOIprefix\doi{10.2139/ssrn.1028206}.
\bibitem[{Fama(1970)}]{fama_1970}
\bibinfo{author}{Fama, E.~F.} (\bibinfo{year}{1970}).
\newblock \bibinfo{title}{Efficient capital markets: A review of theory and
  empirical work}.
\newblock {\it \bibinfo{journal}{The Journal of Finance}\/},  {\it
  \bibinfo{volume}{25}\/}, \bibinfo{pages}{383--417}.
\bibitem[{Garman \& Klass(1980)}]{garman_1980}
\bibinfo{author}{Garman, M.~B.}, \& \bibinfo{author}{Klass, M.~J.}
  (\bibinfo{year}{1980}).
\newblock \bibinfo{title}{On the estimation of security price volatilities from
  historical data}.
\newblock {\it \bibinfo{journal}{The Journal of Business}\/},  {\it
  \bibinfo{volume}{53}\/}, \bibinfo{pages}{67--78}.
\bibitem[{Geczy \& Samonov(2016)}]{geczy_2016}
\bibinfo{author}{Geczy, C.~C.}, \& \bibinfo{author}{Samonov, M.}
  (\bibinfo{year}{2016}).
\newblock \bibinfo{title}{Two centuries of price-return momentum}.
\newblock {\it \bibinfo{journal}{Financial Analysts Journal}\/},  {\it
  \bibinfo{volume}{72}\/}, \bibinfo{pages}{32--56}.
\bibitem[{Georgopoulou \& Wang(2016)}]{georgopoulou_wang_2016}
\bibinfo{author}{Georgopoulou, A.~A.}, \& \bibinfo{author}{Wang, G.~J.}
  (\bibinfo{year}{2016}).
\newblock \bibinfo{title}{The trend is your friend: Time-series momentum
  strategies across equity and commodity markets}.
\newblock {\it \bibinfo{journal}{SSRN Electronic Journal}\/}, .
\bibitem[{Ghosn \& Bengio(1996)}]{ghosn_1996}
\bibinfo{author}{Ghosn, J.}, \& \bibinfo{author}{Bengio, Y.}
  (\bibinfo{year}{1996}).
\newblock \bibinfo{title}{Multi-task learning for stock selection}.
\newblock In \bibinfo{editor}{M.~Mozer}, \bibinfo{editor}{M.~Jordan}, \&
  \bibinfo{editor}{T.~Petsche} (Eds.), {\it \bibinfo{booktitle}{Advances in
  Neural Information Processing Systems}\/}.
\newblock \bibinfo{publisher}{MIT Press} volume~\bibinfo{volume}{9}.
\newblock \URLprefix
  \url{https://proceedings.neurips.cc/paper_files/paper/1996/file/1d72310edc006dadf2190caad5802983-Paper.pdf}.
\bibitem[{Graham(2020)}]{robertson_management_2020}
\bibinfo{author}{Graham, R.} (\bibinfo{year}{2020}).
\newblock \bibinfo{title}{We see risk where others may not: 60/40 in 2020
  vision}.
\newblock \URLprefix
  \url{https://www.man.com/maninstitute/60-40-in-2020-vision}.
\bibitem[{Harvey et~al.(2021)Harvey, Rattray \&
  Hemert}]{harvey_rattray_hemert_2021}
\bibinfo{author}{Harvey, C.~R.}, \bibinfo{author}{Rattray, S.}, \&
  \bibinfo{author}{Hemert, O.~v.} (\bibinfo{year}{2021}).
\newblock {\it \bibinfo{title}{Strategic risk management: Designing portfolios
  and managing risk}\/}.
\newblock \bibinfo{publisher}{John Wiley \&amp; Sons, Inc.}
\bibitem[{Hochreiter \& Schmidhuber(1997)}]{schmidhuber_1997}
\bibinfo{author}{Hochreiter, S.}, \& \bibinfo{author}{Schmidhuber, J.}
  (\bibinfo{year}{1997}).
\newblock \bibinfo{title}{{Long Short-Term Memory}}.
\newblock {\it \bibinfo{journal}{Neural Computation}\/},  {\it
  \bibinfo{volume}{9}\/}, \bibinfo{pages}{1735--1780}.
  \DOIprefix\doi{10.1162/neco.1997.9.8.1735}.
\bibitem[{Hurst et~al.(2017)Hurst, Ooi \& Pedersen}]{hurst_2017}
\bibinfo{author}{Hurst, B.}, \bibinfo{author}{Ooi, Y.~H.}, \&
  \bibinfo{author}{Pedersen, L.~H.} (\bibinfo{year}{2017}).
\newblock \bibinfo{title}{A century of evidence on trend-following investing}.
\newblock {\it \bibinfo{journal}{The Journal of Portfolio Management}\/},  {\it
  \bibinfo{volume}{44}\/}, \bibinfo{pages}{15--29}.
\bibitem[{Jegadeesh \& Titman(1993)}]{jegadeesh_titman_1993}
\bibinfo{author}{Jegadeesh, N.}, \& \bibinfo{author}{Titman, S.}
  (\bibinfo{year}{1993}).
\newblock \bibinfo{title}{Returns to buying winners and selling losers:
  Implications for stock market efficiency}.
\newblock {\it \bibinfo{journal}{The Journal of Finance}\/},  {\it
  \bibinfo{volume}{48}\/}, \bibinfo{pages}{65--91}.
\bibitem[{Jegadeesh \& Titman(2001)}]{jegadeesh_titman_2001}
\bibinfo{author}{Jegadeesh, N.}, \& \bibinfo{author}{Titman, S.}
  (\bibinfo{year}{2001}).
\newblock \bibinfo{title}{Profitability of momentum strategies: An evaluation
  of alternative explanations}.
\newblock {\it \bibinfo{journal}{The Journal of Finance}\/},  {\it
  \bibinfo{volume}{56}\/}, \bibinfo{pages}{699--720}.
\bibitem[{Kang et~al.(2022)Kang, Chen, Jia, Wei, Deng \& Qian}]{yanzhe_2022}
\bibinfo{author}{Kang, Y.}, \bibinfo{author}{Chen, L.}, \bibinfo{author}{Jia,
  N.}, \bibinfo{author}{Wei, W.}, \bibinfo{author}{Deng, J.}, \&
  \bibinfo{author}{Qian, H.} (\bibinfo{year}{2022}).
\newblock \bibinfo{title}{A cwgan-gp-based multi-task learning model for
  consumer credit scoring}.
\newblock {\it \bibinfo{journal}{Expert Systems with Applications}\/},  {\it
  \bibinfo{volume}{206}\/}, \bibinfo{pages}{117650}. \URLprefix
  \url{https://www.sciencedirect.com/science/article/pii/S0957417422009538}.
  \DOIprefix\doi{https://doi.org/10.1016/j.eswa.2022.117650}.
\bibitem[{Kim(2003)}]{kim_2003}
\bibinfo{author}{Kim, K.-J.} (\bibinfo{year}{2003}).
\newblock \bibinfo{title}{Financial time series forecasting using support
  vector machines}.
\newblock {\it \bibinfo{journal}{Neurocomputing}\/},  {\it
  \bibinfo{volume}{55}\/}, \bibinfo{pages}{307--319}.
  \DOIprefix\doi{10.1016/S0925-2312(03)00372-2}.
\newblock \bibinfo{note}{Support Vector Machines}.
\bibitem[{Kishore et~al.(2008)Kishore, Brandt, Santa-Clara \&
  Venkatachalam}]{kishore_2008}
\bibinfo{author}{Kishore, R.}, \bibinfo{author}{Brandt, M.},
  \bibinfo{author}{Santa-Clara, P.}, \& \bibinfo{author}{Venkatachalam, M.}
  (\bibinfo{year}{2008}).
\newblock \bibinfo{title}{Earnings announcements are full of surprises}.
\newblock {\it \bibinfo{journal}{SSRN Electronic Journal}\/}, .
\bibitem[{Law \& Shawe-Taylor(2017)}]{law_2017}
\bibinfo{author}{Law, T.}, \& \bibinfo{author}{Shawe-Taylor, J.}
  (\bibinfo{year}{2017}).
\newblock \bibinfo{title}{Practical bayesian support vector regression for
  financial time series prediction and market condition change detection}.
\newblock {\it \bibinfo{journal}{Quantitative Finance}\/},  {\it
  \bibinfo{volume}{17}\/}, \bibinfo{pages}{1--14}.
  \DOIprefix\doi{10.1080/14697688.2016.1267868}.
\bibitem[{Levine \& Pedersen(2016)}]{levine_2016}
\bibinfo{author}{Levine, A.}, \& \bibinfo{author}{Pedersen, L.~H.}
  (\bibinfo{year}{2016}).
\newblock \bibinfo{title}{Which trend is your friend?}
\newblock {\it \bibinfo{journal}{Financial Analysts Journal}\/},  {\it
  \bibinfo{volume}{72}\/}, \bibinfo{pages}{51--66}.
\bibitem[{Liebel \& K{\"{o}}rner(2018)}]{lukas_2018}
\bibinfo{author}{Liebel, L.}, \& \bibinfo{author}{K{\"{o}}rner, M.}
  (\bibinfo{year}{2018}).
\newblock \bibinfo{title}{Auxiliary tasks in multi-task learning}.
\newblock {\it \bibinfo{journal}{CoRR}\/},  {\it
  \bibinfo{volume}{abs/1805.06334}\/}.
  \href{http://arxiv.org/abs/1805.06334}{\tt arXiv:1805.06334}.
\bibitem[{Lim et~al.(2019)Lim, Zohren \& Roberts}]{bryan_2019}
\bibinfo{author}{Lim, B.}, \bibinfo{author}{Zohren, S.}, \&
  \bibinfo{author}{Roberts, S.} (\bibinfo{year}{2019}).
\newblock \bibinfo{title}{Enhancing time series momentum strategies using deep
  neural networks}.
\newblock {\it \bibinfo{journal}{arXiv preprint arXiv:1904.04912}\/}, .
  \DOIprefix\doi{10.48550/ARXIV.1904.04912}.
\bibitem[{Ma \& Tan(2022)}]{tao_ma_2022}
\bibinfo{author}{Ma, T.}, \& \bibinfo{author}{Tan, Y.} (\bibinfo{year}{2022}).
\newblock \bibinfo{title}{Stock ranking with multi-task learning}.
\newblock {\it \bibinfo{journal}{Expert Systems with Applications}\/},  {\it
  \bibinfo{volume}{199}\/}, \bibinfo{pages}{116886}. \URLprefix
  \url{https://www.sciencedirect.com/science/article/pii/S0957417422003293}.
  \DOIprefix\doi{https://doi.org/10.1016/j.eswa.2022.116886}.
\bibitem[{Markowitz(1952)}]{harry_1952}
\bibinfo{author}{Markowitz, H.} (\bibinfo{year}{1952}).
\newblock \bibinfo{title}{Portfolio selection}.
\newblock {\it \bibinfo{journal}{The Journal of Finance}\/},  {\it
  \bibinfo{volume}{7}\/}, \bibinfo{pages}{77--91}.
\bibitem[{Mazur et~al.(2021)Mazur, Dang \& Vega}]{mazur_2021}
\bibinfo{author}{Mazur, M.}, \bibinfo{author}{Dang, M.}, \&
  \bibinfo{author}{Vega, M.} (\bibinfo{year}{2021}).
\newblock \bibinfo{title}{Covid-19 and the march 2020 stock market crash.
  evidence from s\&p1500}.
\newblock {\it \bibinfo{journal}{Finance Research Letters}\/},  {\it
  \bibinfo{volume}{38}\/}, \bibinfo{pages}{101690}. \URLprefix
  \url{https://www.sciencedirect.com/science/article/pii/S1544612320306668}.
  \DOIprefix\doi{https://doi.org/10.1016/j.frl.2020.101690}.
\bibitem[{Moskowitz et~al.(2012)Moskowitz, Ooi \& Pedersen}]{moskowitz_2012}
\bibinfo{author}{Moskowitz, T.~J.}, \bibinfo{author}{Ooi, Y.~H.}, \&
  \bibinfo{author}{Pedersen, L.~H.} (\bibinfo{year}{2012}).
\newblock \bibinfo{title}{Time series momentum}.
\newblock {\it \bibinfo{journal}{Journal of Financial Economics}\/},  {\it
  \bibinfo{volume}{104}\/}, \bibinfo{pages}{228--250}.
\bibitem[{Nunes et~al.(2019)Nunes, Gerding, McGroarty \&
  Niranjan}]{manuel_nunes_2019}
\bibinfo{author}{Nunes, M.}, \bibinfo{author}{Gerding, E.},
  \bibinfo{author}{McGroarty, F.}, \& \bibinfo{author}{Niranjan, M.}
  (\bibinfo{year}{2019}).
\newblock \bibinfo{title}{A comparison of multitask and single task learning
  with artificial neural networks for yield curve forecasting}.
\newblock {\it \bibinfo{journal}{Expert Systems with Applications}\/},  {\it
  \bibinfo{volume}{119}\/}, \bibinfo{pages}{362--375}. \URLprefix
  \url{https://www.sciencedirect.com/science/article/pii/S0957417418307322}.
  \DOIprefix\doi{https://doi.org/10.1016/j.eswa.2018.11.012}.
\bibitem[{Parkinson(1980)}]{parkinson_1980}
\bibinfo{author}{Parkinson, M.} (\bibinfo{year}{1980}).
\newblock \bibinfo{title}{The extreme value method for estimating the variance
  of the rate of return}.
\newblock {\it \bibinfo{journal}{The Journal of Business}\/},  {\it
  \bibinfo{volume}{53}\/}, \bibinfo{pages}{61--65}.
\bibitem[{Robinson \& Langley(2017)}]{robinson_langley_2017}
\bibinfo{author}{Robinson, J.}, \& \bibinfo{author}{Langley, B.}
  (\bibinfo{year}{2017}).
\newblock \bibinfo{title}{The 60/40 problem: Examining the traditional 60/40
  portfolio in an uncertain rate environment}.
\newblock \URLprefix
  \url{https://papers.ssrn.com/sol3/papers.cfm?abstract_id=2959015}.
\bibitem[{Rogers \& Satchell(1991)}]{rogers_1991}
\bibinfo{author}{Rogers, L. C.~G.}, \& \bibinfo{author}{Satchell, S.~E.}
  (\bibinfo{year}{1991}).
\newblock \bibinfo{title}{Estimating variance from high, low and closing
  prices}.
\newblock {\it \bibinfo{journal}{The Annals of Applied Probability}\/},  {\it
  \bibinfo{volume}{1}\/}, \bibinfo{pages}{504--512}.
\bibitem[{Rossi(2015)}]{rossi_2015}
\bibinfo{author}{Rossi, M.} (\bibinfo{year}{2015}).
\newblock \bibinfo{title}{The efficient market hypothesis and calendar
  anomalies: a literature review}.
\newblock {\it \bibinfo{journal}{International Journal of Managerial and
  Financial Accounting}\/},  {\it \bibinfo{volume}{7}\/}, \bibinfo{pages}{285}.
\bibitem[{Ruder(2017)}]{sebastian_2017}
\bibinfo{author}{Ruder, S.} (\bibinfo{year}{2017}).
\newblock \bibinfo{title}{An overview of multi-task learning in deep neural
  networks}.
\newblock {\it \bibinfo{journal}{CoRR}\/},  {\it
  \bibinfo{volume}{abs/1706.05098}\/}.
  \href{http://arxiv.org/abs/1706.05098}{\tt arXiv:1706.05098}.
\bibitem[{Sapankevych \& Sankar(2009)}]{sapankevych_2009}
\bibinfo{author}{Sapankevych, N.~I.}, \& \bibinfo{author}{Sankar, R.}
  (\bibinfo{year}{2009}).
\newblock \bibinfo{title}{Time series prediction using support vector machines:
  A survey}.
\newblock {\it \bibinfo{journal}{IEEE Computational Intelligence Magazine}\/},
  {\it \bibinfo{volume}{4}\/}.
\bibitem[{Sebastian~Thrun(1998)}]{thrun_1998}
\bibinfo{author}{Sebastian~Thrun, L.~P.} (\bibinfo{year}{1998}).
\newblock \bibinfo{title}{Learning to learn: Introduction and overview}.
\newblock {\it \bibinfo{journal}{Learning to Learn}\/},  (pp.
  \bibinfo{pages}{3--17}).
\bibitem[{Sharpe(1994)}]{sharpe_1994}
\bibinfo{author}{Sharpe, W.~F.} (\bibinfo{year}{1994}).
\newblock \bibinfo{title}{The sharpe ratio}.
\newblock {\it \bibinfo{journal}{The Journal of Portfolio Management}\/},  {\it
  \bibinfo{volume}{21}\/}, \bibinfo{pages}{49--58}.
\bibitem[{{Stevens Analytics}(2020)}]{steven_analytics}
\bibinfo{author}{{Stevens Analytics}} (\bibinfo{year}{2020}).
\newblock \bibinfo{title}{Trend: Stevens analytics. continuous futures: Future
  daily open price | symbol: -- | exchange: All exchanges | symbol: Ed,
  09/28/1983 - 10/23/2020. data planet™ statistical datasets: A sage
  publishing resource dataset-id: 095-001-006}.
\newblock \DOIprefix\doi{10.6068/DP17560842C1658}.
\bibitem[{Tay \& Cao(2001)}]{francis_2001}
\bibinfo{author}{Tay, F.~E.}, \& \bibinfo{author}{Cao, L.}
  (\bibinfo{year}{2001}).
\newblock \bibinfo{title}{Application of support vector machines in financial
  time series forecasting}.
\newblock {\it \bibinfo{journal}{Omega}\/},  {\it \bibinfo{volume}{29}\/},
  \bibinfo{pages}{309--317}. \DOIprefix\doi{10.1016/S0305-0483(01)00026-3}.
\bibitem[{Vaswani et~al.(2017)Vaswani, Shazeer, Parmar, Uszkoreit, Jones,
  Gomez, Kaiser \& Polosukhin}]{vaswani_2017}
\bibinfo{author}{Vaswani, A.}, \bibinfo{author}{Shazeer, N.},
  \bibinfo{author}{Parmar, N.}, \bibinfo{author}{Uszkoreit, J.},
  \bibinfo{author}{Jones, L.}, \bibinfo{author}{Gomez, A.~N.},
  \bibinfo{author}{Kaiser, L.~u.}, \& \bibinfo{author}{Polosukhin, I.}
  (\bibinfo{year}{2017}).
\newblock \bibinfo{title}{Attention is all you need}.
\newblock In \bibinfo{editor}{I.~Guyon}, \bibinfo{editor}{U.~V. Luxburg},
  \bibinfo{editor}{S.~Bengio}, \bibinfo{editor}{H.~Wallach},
  \bibinfo{editor}{R.~Fergus}, \bibinfo{editor}{S.~Vishwanathan}, \&
  \bibinfo{editor}{R.~Garnett} (Eds.), {\it \bibinfo{booktitle}{Advances in
  Neural Information Processing Systems}\/}.
\newblock \bibinfo{publisher}{Curran Associates, Inc.}
  volume~\bibinfo{volume}{30}.
\bibitem[{Wood et~al.(2021)Wood, Giegerich, Roberts \& Zohren}]{kieran_2021}
\bibinfo{author}{Wood, K.}, \bibinfo{author}{Giegerich, S.},
  \bibinfo{author}{Roberts, S.}, \& \bibinfo{author}{Zohren, S.}
  (\bibinfo{year}{2021}).
\newblock \bibinfo{title}{Trading with the momentum transformer: An intelligent
  and interpretable architecture}.
\newblock \DOIprefix\doi{10.48550/ARXIV.2112.08534}.
\bibitem[{Wood et~al.(2022)Wood, Giegerich, Roberts \&
  Zohren}]{wood_giegerich_roberts_zohren_2022}
\bibinfo{author}{Wood, K.}, \bibinfo{author}{Giegerich, S.},
  \bibinfo{author}{Roberts, S.}, \& \bibinfo{author}{Zohren, S.}
  (\bibinfo{year}{2022}).
\newblock \bibinfo{title}{Trading with the momentum transformer: An intelligent
  and interpretable architecture}.
\newblock \URLprefix \url{https://arxiv.org/abs/2112.08534}.
\bibitem[{Yang \& Zhang(2000)}]{yang_2000}
\bibinfo{author}{Yang, D.}, \& \bibinfo{author}{Zhang, Q.}
  (\bibinfo{year}{2000}).
\newblock \bibinfo{title}{Drift‐independent volatility estimation based on
  high, low, open, and close prices}.
\newblock {\it \bibinfo{journal}{The Journal of Business}\/},  {\it
  \bibinfo{volume}{73}\/}, \bibinfo{pages}{477--492}.
\bibitem[{Yu et~al.(2022)Yu, Wynter \& Lim}]{yu_2022}
\bibinfo{author}{Yu, P.}, \bibinfo{author}{Wynter, L.}, \&
  \bibinfo{author}{Lim, S.~H.} (\bibinfo{year}{2022}).
\newblock \bibinfo{title}{Federated reinforcement learning for portfolio
  management}.
\newblock In \bibinfo{editor}{H.~Ludwig}, \& \bibinfo{editor}{N.~Baracaldo}
  (Eds.), {\it \bibinfo{booktitle}{Federated Learning: A Comprehensive Overview
  of Methods and Applications}\/} (pp. \bibinfo{pages}{467--482}).
\newblock \bibinfo{address}{Cham}: \bibinfo{publisher}{Springer International
  Publishing}.
\newblock \URLprefix \url{https://doi.org/10.1007/978-3-030-96896-0_21}.
  \DOIprefix\doi{10.1007/978-3-030-96896-0_21}.
\bibitem[{Yuan et~al.(2023)Yuan, Ma, Wang, Zhang \& Li}]{chenxun_2023}
\bibinfo{author}{Yuan, C.}, \bibinfo{author}{Ma, X.}, \bibinfo{author}{Wang,
  H.}, \bibinfo{author}{Zhang, C.}, \& \bibinfo{author}{Li, X.}
  (\bibinfo{year}{2023}).
\newblock \bibinfo{title}{Covid19-mlsf: A multi-task learning-based stock
  market forecasting framework during the covid-19 pandemic}.
\newblock {\it \bibinfo{journal}{Expert Systems with Applications}\/},  {\it
  \bibinfo{volume}{217}\/}, \bibinfo{pages}{119549}. \URLprefix
  \url{https://www.sciencedirect.com/science/article/pii/S0957417423000507}.
  \DOIprefix\doi{https://doi.org/10.1016/j.eswa.2023.119549}.
\bibitem[{Zhang et~al.(2020)Zhang, Zohren \&
  Roberts}]{zhang_zohren_roberts_2020}
\bibinfo{author}{Zhang, Z.}, \bibinfo{author}{Zohren, S.}, \&
  \bibinfo{author}{Roberts, S.} (\bibinfo{year}{2020}).
\newblock \bibinfo{title}{Deep learning for portfolio optimisation}.
\newblock \URLprefix
  \url{https://papers.ssrn.com/sol3/papers.cfm?abstract_id=3613600}.
\bibitem[{Zou \& Herremans(2022)}]{zou2022multimodal}
\bibinfo{author}{Zou, Y.}, \& \bibinfo{author}{Herremans, D.}
  (\bibinfo{year}{2022}).
\newblock \bibinfo{title}{A multimodal model with twitter finbert embeddings
  for extreme price movement prediction of bitcoin}.
\newblock {\it \bibinfo{journal}{arXiv preprint arXiv:2206.00648}\/}, .

\end{thebibliography}

\end{document}